\documentstyle[12pt]{article}
\begin{document}
\title{
\begin{flushright}
  {\normalsize Wash.\ U.\ HEP/93-33} \\
[0.5em]
{\normalsize hep-lat/9307001}
\\[2.5em]
\end{flushright}
A Lattice Study of the Gluon Propagator in Momentum Space} 

\author{C. Bernard$^{(1)}$, C. Parrinello$^{(2,3)}$ and
        A. Soni$^{(3)}$
\\[1.5em]
$^{(1)}$: Department of Physics, Washington University\\
St. Louis, MO 63130, USA\\[0.2em]
$^{(2)}$: Physics Department, New York University\\
4 Washington Place, New York, NY 10003, USA\\[0.2em]
$^{(3)}$: Physics Department, Brookhaven National Laboratory\\
Upton, NY 11973, USA\\[0.2em]
}
\maketitle
\newpage

\begin{abstract}
We consider 
pure glue QCD 
at $\beta=5.7, \ \beta=6.0 $ and $\beta=6.3$.
We evaluate the 
gluon propagator both in time at zero 3-momentum 
and in momentum space.
\null From the former quantity we obtain evidence for a dynamically generated 
effective mass,
which at $\beta=6.0$ and $\beta=6.3$ increases with the time separation of 
the sources, in agreement with 
earlier results.
The momentum space propagator $G(k)$ provides further evidence for 
mass generation. 
In particular, at $\beta=6.0$,
for $k \leq 1$ GeV,
 the propagator $G(k)$
can be fit to
a continuum formula proposed by Gribov
and others, which contains a mass scale $b$,
presumably related to the hadronization mass scale.
For higher momenta
Gribov's model no longer
provides a good fit,
as $G(k)$ tends rather
to follow an inverse power law $\approx 1 / k^{2 + \gamma}$.
The results at $\beta=6.3$ are consistent with those at $\beta=6.0$,
but only the high momentum region is accessible on this lattice.
We find $b$ in the range of three to four hundred MeV and
$\gamma$ about 0.7.
Fits to particle $+$ ghost expressions are also possible, often resulting in 
low values for $\chi^2_{dof}$, but the parameters are very poorly determined. 
On the other hand, at $\beta=5.7$ (where we can only study momenta up to 
1 GeV) $G(k)$ is best fit to a simple massive boson propagator with mass $m$. 
We argue that such a discrepancy may be related to a lack of scaling 
for low momenta
at $\beta=5.7$. 
 
\null From our results, the study of correlation functions in momentum space 
looks 
promising,
especially
 because the data points in Fourier space turn out to be 
much less correlated than in real space.
\end{abstract}

\maketitle
\section{Introduction}
The possibility of studying nonperturbatively on the lattice gauge-dependent 
quantities provides in principle a unique tool to test QCD at the level of the
basic fields entering the continuum Lagrangian. From this point of view, the
gluon propagator in the pure glue theory is perhaps the simplest 
quantity. From its study one expects to obtain among other things a better 
understanding of the infrared behavior of the theory and of the mechanism of 
gluon confinement. 
As a result we may also hope to acquire a better understanding of the 
hadronization phenomena and of the glueball spectrum \cite{cns}.

Let us first review what is known from perturbation theory: consider the 
expression
(in Minkowski space-time)
\begin{equation}
D_{\mu \nu}^{a b} (k) \equiv - i \int d^4 x \ \langle {\rm O} | \ T [A_{\mu}^a 
(x) \ A_{\nu}^b (0) ] \ | {\rm O} \rangle \ e^{i k \cdot x}
\end{equation}
\null From the Faddeev-Popov quantization in a class of covariant 
gauges one gets the
simple Slavnov identity 
\begin{equation}
k^{\mu} k^{\nu} \ D_{\mu \nu}^{a b} (k) = - i \alpha \ \delta^{a b}
\label{eq:Sla}
\end{equation}
where $\alpha$ is the gauge parameter. From Lorentz covariance, the general 
solution to (\ref{eq:Sla}) can be written as
\begin{equation}
D_{\mu \nu}^{a b} (k) = - i \delta^{a b} \left[ \left( g_{\mu \nu} - 
{k_{\mu} k_{\nu} \over k^2} \right) {1 \over 1 + \Pi (k^2, \alpha)} + 
\alpha {k_{\mu} k_{\nu} \over k^2} \right] {1 \over k^2}
\end{equation}
Eq.\ (\ref{eq:Sla}) implies that the longitudinal part of the propagator gets 
trivially renormalized, so that the vacuum polarization $\Pi (k^2, \alpha)$
just renormalizes the transverse propagator. The renormalization constant $Z_3$
for the gauge fields, $A_i = Z_3^{1/2} A^R_i$ is defined as
\begin{equation}
Z_3^{-1} = 1 + \Pi ({\Lambda \over \mu}, \alpha)
\end{equation}
where $\Lambda$ is an ultraviolet cut-off and $k^2 = - \mu^2$ is a spacelike 
value of the momentum. In general $Z_3$ is gauge dependent.

One can then rewrite the unrenormalized transverse gluon propagator as
\begin{equation}
D_{\mu \nu}^{T \ a b} (k^2) \vert_{k^2 = - \mu^2} = {i \over \mu^2} Z_{3} 
\left( g_{\mu \nu} + {k_{\mu} k_{\nu} \over \mu^2} \right) \delta^{a b}
\end{equation} 

Unlike QED, in QCD we know from the properties of the $\beta$-function 
that perturbation theory can only be applied to the study of the large-$k^2$
behavior of Green's functions. In such a region one gets at the 1-loop level 
\cite{Cheng}
\begin{equation}
Z_3 = 1 + {g^2 \over 16 \pi^2} \left({13 \over 3} - \alpha \right) \cdot 3 
\ ln ({\Lambda \over \mu})
\end{equation}
 In general, 
at each order in 
$g^2$ perturbative corrections depend 
logarithmically in $k$, so that
in the deep Euclidean region ($k^2 \rightarrow \infty$)
gluon propagators behave essentially in the same way
as the photon propagators,
as long as one considers a finite order in 
perturbation theory.

In the infrared region  things are probably 
different. 
In fact, in the simpler QED case perturbation theory is still 
reliable, and one can evaluate order by order the vacuum polarization function.
In particular, one finds that $\Pi^{QED} (k^2 = 0) $ is finite 
if all the fermions are 
massive (we omit from now on 
the explicit $\alpha-$dependence of 
vacuum polarization functions). This implies that the $k^2 = 0$ pole of the 
free photon propagator is still present after radiative corrections are taken 
into account, so that the photon remains a massless particle. Such corrections 
only affect the residue at the pole, resulting in charge renormalization.

On the other hand, consider QCD without quarks. The corresponding Lagrangian 
does not contain any mass scale in 4 dimensions, yet if the theory is  
confining a mass scale must be dynamically generated in some way, since the 
confinement potential $V(r) = K r$ contains such a scale.
Such a mass $M$ cannot be generated in perturbation theory,
since it must satisfy
\begin{equation}
M(g, \mu) = \mu \ exp \left[ - \int^g {dg^{'} \over \beta(g^{'})} \right]
\end{equation}
where $\mu$ is a renormalization scale $\mu$.
For small $g$, one has
\begin{equation}
M(g, \mu) \approx \mu \ exp \left[ - {{\rm const.} \over g^2} \right] \qquad 
{\rm when} 
\ g \rightarrow 0
\end{equation}
so that $M(g)$ has an essential singularity at $g=0$.

Such a mass scale may show up in the vacuum polarization function for the 
gluon. Indeed, while contributions to $\Pi (k^2)$ proportional to finite 
powers of $g$ have for large $k^2$ a logarithmic momentum dependence, 
non-perturbative effects may generate terms in $\Pi (k^2)$ proportional to 
{\it negative} powers of the squared 4-momentum,
for instance 
\begin{equation}
\Pi (k^2) = - {m^2 (g, \mu) \over k^2} + {b^4 (g, \mu) \over k^4} + O(g^2)
\label{eq:pola}
\end{equation}
where $m(g, \mu), \ b (g, \mu)$ 
have the dimension of a mass and depend nonanalytically on $g$.
For example, the case $b = 0, \ m \not= 0$ gives rise to a mass pole 
in the gluon propagator and corresponds to 
the standard Schwinger mechanism \cite{Sch}.

In the above formula $O(g^2)$ denotes 
contributions which can be represented as a power series in $g^2$. 
Such series, when truncated to a finite order in $g^2$, behave like 
polynomials in $ln (k^{2} / \mu^{2})$ for large $k^2$.
On the other hand, the sum of the contributions to all orders in 
$g^2$
can generate an anomalous dimension $\gamma$ 
\cite{Cheng2}; in such a case  
for $k^{2} \rightarrow \infty$ the propagator behaves like 
$1 / k^{2 + \gamma}$. 
The nonperturbative behavior of the Euclidean gluon propagator has been
investigated  in the continuum by many authors with different methods and in
different gauges  \cite{Gri,Cornwall,Stingl,Zwa,Nami,Mandel}. 
In some of these attempts 
a very singular gluon propagator was found, behaving like $k^{-4}$ in the 
limit $k^2 \rightarrow 0$ \cite{Mandel},
 and a confining property was inferred 
from such behavior. On the 
other hand, other work points towards the elimination
at the non-perturbative level of the singularity at $k^2=0$ of the 
propagator \cite{Cornwall,Stingl,Nami}, 
as a consequence of dynamical mass generation.

In particular, a very peculiar momentum dependence, consistent with the above
scenario, has been predicted  as 
arising from a modification of the standard path integral Faddeev-Popov 
formula in the Landau gauge by the introduction of a 
nonperturbative
gauge-fixing procedure \cite{Gri,Zwa}. Such improved implementation of 
the Landau gauge is expressed by the equations 
\begin{equation}
\partial \cdot A = 0 \qquad \qquad {\rm and} \qquad \qquad FP [A] > 0
\label{eq:gauge}
\end{equation} 
where $FP [A]$ is the Faddeev-Popov operator in the Landau gauge, which in 
general is not positive definite. The positivity requirement in 
(\ref{eq:gauge}) can be seen as a recipe to get rid (although not completely 
\cite{DAZW})
of Gribov copies \cite{Gri}. In the gauge 
(\ref{eq:gauge}), the (transverse) gluon propagator in momentum space has been 
argued to be of the form \cite{Gri,Zwa} 
\begin{equation}
G (k) \approx {k^2 \over k^4 + b^4}
\protect\label{eq:prop}
\end{equation}
where $b$ is a dynamically generated mass scale. 
Eq.\ (\protect\ref{eq:prop}) corresponds to the case $m =0, \ b \not= 0$ 
in (\ref{eq:pola}). 
It implies that in 
the continuum  
\begin{equation}
G (\vec{k}=0,t) \approx e^{- {b \over \sqrt{2}} t} \ \left( 
cos({b \over \sqrt{2}} t) - sin({b \over \sqrt{2}} t) \right)
\protect\label{eq:propspace}
\end{equation}
Remarkably, the same predictions were also obtained in the study of
Schwinger-Dyson equations \cite{Stingl}.

The above expression lends itself to intriguing speculations: the absence of 
any particle singularity on the real $k^2$ axis predicts the absence of 
an asymptotic gluon state. It may describe a short-lived excitation, giving 
rise to a gluon jet. In this framework, the mass scale $b$ appearing in 
the above formulas may perhaps be interpreted as a hadronization scale.

 The lattice gluon field can be defined as \cite{Mand}:
\begin{equation}
A_{\mu} (n) \equiv 
{U_{\mu} (n) - U_{\mu}^{\dagger} (n) 
\over 2 i a} - \frac{1}{3} 
Tr \left(  
{U_{\mu} (n) - U_{\mu}^{\dagger} (n) 
\over 2 i a}  \right)
\label{eq:gluone}
\end{equation}
where $a$ is the lattice spacing. Thus the lattice gluon propagator in 
$x$-space is the expectation value of:
\begin{equation}
G_{\mu \nu} (x, y) \equiv Tr \left(A_{\mu} (x) \ A_{\nu} (y) \right) 
\label{eq:prolat}
\end{equation}

An important point is that on the lattice one can define and implement the
analogue of the gauge condition (\ref{eq:gauge}).
In fact, given any link configuration $\{ U \}$, one can define a function of
the gauge transformations $g$ on $\{ U \}$ 
\begin{equation} F_U [g] \equiv - {1 \over V} \sum_{n, \mu} \ Re \ Tr \ \left(
U_{\mu}^{g} (n) + U_{\mu}^{g \dagger} (n- \hat{\mu}) \right), \label{eq:effecl}
\end{equation} 
where $V$ is the lattice volume and $U^{g}$ indicates the gauge-transformed 
link
$U_{\mu}^{g} (n) \equiv g (n) U_{\mu} (n) g^{\dagger} (n + \hat{\mu})$. An
iterative minimization of $F_U [g]$ obtained by performing suitable gauge
transformations generates a configuration $\{ U^{\bar{g}} \}$ 
which satisfies the lattice version of (\ref{eq:gauge}), defined in terms of a 
lattice Faddeev-Popov operator \cite{Zwavan}. This is just the Hessian matrix 
associated
with
$F_{U} [g]$, 
that is,
 ${\delta^{2} F_{U} [g] \over \delta g_{1} \ 
\delta g_{2}}$.
As we have already mentioned for the continuum, such a gauge is not 
completely free of Gribov copies. This is also true on the lattice, and 
corresponds to the fact that for a fixed configuration $\{ U \}$ the 
function (\ref{eq:effecl}) may have several local minima (see, for example, 
\cite{Mari}). 

In general, Zwanziger \cite{Zwavan} showed that on the lattice 
one has qualitatively the 
same scenario for the Landau gauge as in the continuum.
Indeed, there exists 
a bounded region $\Omega$, defined by the positivity requirement for the 
lattice Faddeev-Popov operator, 
which satisfies bounds analogous to the ones derived for the continuum model. 
Considering then a restriction of the functional integration to the region 
$\Omega$, 
Zwanziger
 was able to obtain predictions for the lattice gluon propagator 
consistent with the continuum ones given in 
(\protect\ref{eq:prop}) and (\protect\ref{eq:propspace}).

At this point it is natural to try and test numerically 
predictions like (\protect\ref{eq:prop}) and (\protect\ref{eq:propspace}).

Numerical studies have been performed in the past years for the zero spatial
momentum Fourier transform of (\ref{eq:prolat}), namely $G(\vec{k}=0,t) \equiv
\sum_{i=1}^{3} \ G_{i i} (\vec{k}=0,t)$ \cite{Mand,Gup,Soni}. These studies
reported  evidence of an effective gluon mass that increases with the time
separation for short time intervals.
This feature, which would be  unacceptable for the propagator of a
real physical particle since it violates the K\"{a}llen-Lehmann representation,
is in qualitative agreement with the continuum prediction 
(\protect\ref{eq:propspace}) 
and may be in principle acceptable for a confined particle \cite{Stingl,Mand}. 
Another lattice approach to the gluon mass was given in  \cite{cb}.

Our work aims to test at a more quantitative level continuum predictions and
to extend the above results through the study of the gluon propagator at 
nonzero momenta. 

\section{Numerical Results}
 We study pure glue QCD on $16^3 \times 40$ and $24^3 \times 40$ lattices
 at $\beta=6.0$,
 on a $24^4$ lattice at $\beta=6.3$ and on a $16^3 \times 24$ lattice at 
$\beta=5.7$.

\subsection{Technical Remarks}

It is worth remarking that, unlike simulations involving quenched 
quark propagators, evaluations of purely gluonic correlation functions can take
full advantage of the translational symmetry of the theory in order to improve
statistics. On the other hand, such quantities turn out to be very sensitive
to the numerical accuracy of gauge fixing. 
Empirically, 
at 
$\beta=6.0$ and $\beta=6.3$ we find that 
when the minimization of $F_U [g]$ has reached 
an accuracy 
such that in $O (50)$ iterations $F_U [g]$ changes less than $\approx .05 \%$,
then the signal for the propagators is 
sufficiently 
stable against additional gauge fixing.
In other words, the variation in each data point for the propagator 
arising from 
additional gauge fixing is typically much smaller than the final error 
bar associated to the data point.
On the contrary, we will see that at $\beta=5.7$ our stability requirement 
for $F_U [g]$ does not suffice to guarantee a completely stable propagator. 

Even at $\beta=6.0$ and $\beta=6.3$, though, 
when our empirical criterion gets satisfied 
the system has not yet reached complete equilibrium, 
in spite of the fact that such an accuracy is roughly 
one order of magnitude better than the typical one adopted in simulations
of hadron phenomenology. 
 This can be seen by performing the following test \cite{Mand}: 
in the Landau gauge $\partial_{\mu} A_{\mu} = 0$ it follows from the periodic 
boundary conditions that 
\begin{equation}
A_{0} (t) \equiv \sum_{\bar{x}} \ A_{0} (\bar{x}, t) 
\end{equation}
should not depend on $t$. In lattice language, this means that once the
Landau gauge has been numerically implemented in a configuration, then the sum
over the sites in a fixed timeslice  of the time component of the gauge field
should be the same on each timeslice. 
In Fig.~\ref{fig:test} we plot one of the diagonal elements of the matrix 
$A_{0}(t) $ as a function of $\ t \ $ for one of our gauge-fixed configurations
on the $16^3 \times 40$ lattice at $\beta=6.0$. 
This test shows that 
even when the accuracy of our numerical gauge fixing is 
sufficient for the gluon propagator, there are other quantities for which the 
gauge fixing 
need not be adequate. 
As a consequence, it would be dangerous in a calculation to rely on some 
standard {\it a priori} criterion when estimating the required 
precision for the 
gauge fixing, since such precision strongly depends on the specific observable 
under consideration. 
We remark that the 
gauge-fixing test provided by $A_{0}(t) $ is in fact more
stringent than the one used in  \cite{Mand}.
\subsection{Results at $\beta=6.0$ and $\beta=6.3$}
Here we give results for a set of 25 configurations of size $16^3 \times 40$ 
at $\beta=6.0$,
a set of 8 configurations of size $24^3 \times 40$ 
at $\beta = 6.0$, and finally a set of 20 configurations of size $24^4$ 
at $\beta=6.3$. 

As a first step we have evaluated $ G (\vec{k}=0,t)$;
our results confirm that this propagator exhibits a massive decay in time,
with an effective mass $a m(t) \equiv ln ({G (\vec{k}=0,t) /  G
(\vec{k}=0,t+a)})$ that increases with $t$
for short times.
In Fig.~\ref{fig:effmass} we plot 
$a m(t)$ versus $t$ with jackknife errors \cite{Gott} 
on the $16^3 \times 40$ lattice. 
Assuming the value of the inverse lattice spacing $a^{-1} = 2.1$ GeV
at $\beta=6.0$ \cite{Labrenz}, the effective gluon mass $m(t)$ 
ranges approximately
between 220 and 870 MeV. 
A similar behavior is observed for the $24^3 \times 40$ lattice,
with the effective mass ranging between 240 and 1200 MeV.   

At this point we have attempted to fit $G(k=0,t)$ to the
continuum form (\protect\ref{eq:propspace}) and to the form commonly referred 
to as particle $+$ ghost, that is 
\begin{equation}
G (\vec{k}=0,t) = C_1 \ {\rm exp}
(-M_1 t) + C_2 \ {\rm exp}(-M_2 t)
\label{eq:ghost}
\end{equation}
 where $C_2$ is constrained to be negative.

As is well known, 
the data points obtained from a Monte Carlo simulation 
are in general statistically correlated; in the present case, the correlated 
data are the values of the propagator $G (\vec{k}=0,t)$ at different 
timeslices.
For this reason, one should perform $\chi^2$ fits by using the definition 
of $\chi^2$ which involves the full covariance matrix \cite{Tous}. 

By inspection of the covariance matrix for $G (\vec{k}=0,t)$, it turns out that
the off-diagonal matrix elements are typically of the same size as the 
diagonal 
ones, which means that our data points are highly correlated in $t$. 
Consequently, $\chi^2$ fits are not well controlled because the 
covariance matrix is nearly singular. 
  Much higher statistics
would be required to get well-behaved fits.

As it is not possible in this case to make fits using the full covariance 
matrix, we are forced to use the naive definition of $\chi^{2}$, where the data
points are simply weighted by their standard error bar. 
Of course, in such an approach $\chi^2$ has no simple relation
to ``goodness of fit.''

In this approximation, it turns out that Gribov's formula 
provides, for small $t$, a good fit to $G (\vec{k}=0,t)$, better than the one 
obtained from the particle $+$ ghost expression.
In fact, 
although
 $\chi^2_{dof}$ is not a reliable indicator of the goodness of 
a fit when the covariance matrix is not taken into account, nonetheless 
the relative $\chi^2_{dof}$ of two fits provides some indication of which fit 
is better. 
On the $16^3 \times 40$ lattice at $\beta=6.0$,
the lowest value for $\chi^2_{dof}$ obtained from a fit to 
Gribov's formula is $\chi^2_{dof}=.18$, while from the 4-parameter fit to 
particle $+$ ghost one gets $\chi^2_{dof}=.34$. 
Moreover, the latter kind of fit 
is much less stable against varying the initial guess for the fit parameters, 
i.e. many 
local
minima for $\chi^2$ can be found.
 Using $a^{-1} = 2.1$ GeV, we obtain for the $b$ parameter in 
(\protect\ref{eq:propspace}) $b = 237 \pm 7$ MeV, $\chi^2_{dof}=.18$, where the 
error on $b$ is a jackknife one.
Of course one does not get good agreement by using a conventional
4-parameter double exponential form, that is if one constrains $C_2$ in 
(\ref{eq:ghost}) to take positive values. Indeed, in this case 
the effective mass 
would always decrease with $t$, in contrast to what is observed.

The statistical difficulties 
mentioned above forbid a complete analysis of $G (\vec{k}=0,t)$; 
in particular, we cannot effectively 
study the large-$t$ region, where according to (\protect\ref{eq:propspace})
the propagator may become negative and then oscillate.

Our analysis of the momentum space propagator $G(k) \equiv \sum_{\mu=1}^4
G_{\mu \mu}(k)$ does not have the same statistical difficulties
as in the case of $G(\vec{k}=0,t)$. 
This quantity is obtained by performing explicitly the lattice Fourier 
transform of the propagator (\ref{eq:prolat}).

In fact, the covariance matrix associated with $G(k)$ 
turns out to be much more ``diagonal" than the one for $G(\vec{k}=0,t)$; 
in other words, the data points are much less correlated in momentum space 
than they are in $t$. 
We find that $G(k)$ is very well determined 
in a significant interval of 
physical momenta, ranging from the lattice infrared cutoff 
in the time direction  
$k_o = {2 \pi / N_t a}$\  
($N_t = 40$ at $\beta=6.0$)
up to 
$k \approx 2$ GeV, 
which is in fact roughly the value of the ultraviolet cutoff $a^{-1}$
(see Fig.~\ref{fig:3_4}). 
As a consequence we have been able to obtain good fits 
to $G(k)$ taking into account correlations. 
Fig.~\ref{fig:3_4} also shows
that the data points from the two lattices
at $\beta=6.0$ are in good agreement
over most of the momentum range.

At this point we attempt to fit $G(k)$ 
to the continuum formula (\protect\ref{eq:prop}) and, for a  
comparison, to a standard massive propagator $G_{mass} (k) = {A/  (k^2 + 
m^2)}$.
The fits are performed including in $\chi^{2}$ the covariance matrix.
Consider first the $16^3 \times 40$ lattice: 
it turns out that for momenta up to $\approx 1$ GeV formula 
(\protect\ref{eq:prop}) fits
 the data well (see Fig.~\protect\ref{fig:5_7}). 
 Indeed, for such a fit we obtain $\chi^2_{dof} = 1.3$
and $b = 341 \pm 12$ MeV, assuming $a^{-1} = 2.1$ GeV. 
The error is given in terms of the parameter covariance 
matrix \cite{Tous}, and we are 
not including the uncertainty in the value of $a^{-1}$.
 
We compare this result to the best fit that one can obtain from the 
standard massive propagator, for which we obtain $\chi^2_{dof} = 
2.9$ 
\footnote{We have also attempted a fit to a form proposed in \cite{Cornwall}.
 Such a form 
in the low momentum region is very similar to a standard massive propagator. 
The standard massive form appears to provide a better fit.}. 

On the other hand, in the range of momenta $1 - 2$ GeV the formula 
(\protect\ref{eq:prop}) and the massive propagator expression fit the data 
very poorly, resulting in $\chi^2_{dof} > 10$. 
A good fit in this range is obtained by assuming an inverse power 
law behavior 
$G(k) = A / k^{2 + \gamma}$ (see Fig.~\protect\ref{fig:pow_uv_1}). 
We obtain 
$\gamma = .7 \pm .2$, $\chi^2_{dof} = 1$. 
The fact that (\protect\ref{eq:prop}) does not fit the high
momentum region well presumably explains the discrepancy
between the value of $b$ obtained from $G(k)$ at low
momenta and that obtained from $G(\vec k=0,t)$, which depends
on the complete range of momenta.

The situation is 
qualitatively similar on the $24^3 \times 40$ lattice. 
However, presumably due to the lower statistics,
there is clear dispersion in the data (see Fig.~\ref{fig:3_4}),
which means that no fit will be particularly good. In fact, up to 1 
GeV the best fit is given by (\protect\ref{eq:prop}) (see 
Fig.~\ref{fig:5_7}),
 and we obtain $b = 333 \pm 12$ MeV, $\chi^2_{dof} = 5.8$. 
Between 1 and 2 GeV 
 we are unable to perform fits with the covariance 
matrix, due to the poor statistics. Without the covariance matrix, 
 an inverse power law with 
$\gamma \approx .6$
again reproduces 
the data well (see 
 Fig.~\ref{fig:pow_uv_1}).

Summarizing, the data for $G(k)$ at $\beta = 6.0$ 
on the $16^3 \times 40$ lattice up to momenta of order 1 GeV
prefer somewhat 
 formula (\protect\ref{eq:prop}), 
which describes the mass generation $\grave{a} \ la$ Gribov,
over a standard massive propagator.
In the momentum range 1 - 2 GeV the propagator is best reproduced by 
the inverse power law behavior 
$G(k) \approx 1 / k^{2 + \gamma}$, where $\gamma$ could be interpreted as 
the anomalous dimension of $G(k)$. 
 Indeed, our results in such a region could also be sensitive to the 
lattice ultraviolet 
cutoff $a^{-1}$.  The fact that the $\beta=6.3$ results (described below)
agree with those at $\beta=6.0$ gives some
indication that such cutoff effects are not overwhelmingly large.
An additional test for lattice artifacts is  described at the
end of the next section.

On the other hand, 
the results from the $24^3 \times 40$ lattice, due 
to the poor statistics, do not provide by themselves a strong confirmation 
for the above picture. 
Still, in the low momentum region the Gribov 
fit works better than a standard massive one and gives a value for $b$ 
which is consistent with the one from the $16^3 \times 40$ lattice,
 but we get a  high value for $\chi^2_{dof}$. Moreover, in the higher momentum 
region  we are unable to use the covariance matrix.

We consider now a set of 20 configurations on a $24^4$ lattice at $\beta=6.3$.

Starting again from the evaluation of 
$G (\vec{k}=0, t)$, we give in 
Fig.~\ref{fig:2effmass} the effective gluon mass with jackknife errors.
Again, the effective mass $a m(t)$ appears to increase with $t$. 
Assuming $a^{-1} = 3.2$ GeV at $\beta=6.3$ \cite{Labrenz}, $m(t)$ ranges 
approximately between 200 and 670 MeV.

It turns out that we have the same statistics problem as 
at $\beta=6.0$: the data for $G (\vec{k}=0, t)$ 
are highly correlated in $t$, 
so that it is impractical to make fits using the full covariance matrix.

Next we consider the propagator $G (k)$ in momentum space.
We again find that correlations between different values of $k$ are much 
smaller than the correlations in $t$, so that the covariance matrix is now 
well 
behaved. 
 The best fit to $G(k)$ is obtained from the inverse power law 
$G(k) = A / k^{2 + \gamma}$  (see Fig.~\ref{fig:bello}). We 
assume $a^{-1} = 3.2$ GeV \cite{Labrenz} and 
get
$\gamma= .68 \pm .08$,  ${\chi}^{2}_{dof} = 1.9$. 

Given the momentum range 
of such a fit, which starts around 1 GeV and goes up to momenta of the order 
$a^{-1} $, such a result is quite consistent with the behavior observed at
$\beta=6.0$. In particular, the value for 
 $\gamma$ agrees with the one obtained on the $16^3 \times 40$ 
lattice. 

Both at $\beta=6.0$ and $\beta=6.3$ we have also attempted 4-parameter fits,
 in terms of the particle $+$ ghost form 
$G(k) = A/(k^2 + m_1^2) + B/(k^2 + m_2^2)$, with $A>0$ and $B<0$. 
Although
low values of ${\chi}^{2}_{dof}$ can sometimes
be obtained, the determination 
of the 
fit parameters is very poor.
 
\subsection{Results at $\beta=5.7$ and Comparison of Lattices}
Here we give results for a set of 16 configurations on a $16^3 \times 24$ 
lattice at $\beta=5.7$. 
After implementing our standard level of gauge fixing, we find that 
the data for 
$G(\vec{k}=0, t)$ are, as usual, not 
suitable for $\chi^2$ fits with the full covariance matrix.
When fit without 
the covariance matrix, 
 the data seem to be best reproduced by a simple decreasing exponential, 
unlike 
what happened on the other data sets at weaker couplings. 
Correspondingly,
there is no clear evidence for an increase of the 
effective gluon mass $a m(t)$ 
on a significant time interval (see Fig.~\ref{fig:effmass57}).

In addition, the results for $G(k)$ on this lattice differ from what we 
observed 
at weaker couplings. Assuming $a^{-1} = 1.2$ GeV, our data for $G(k)$ 
cover a momentum range up to roughly 1 GeV.
 The best fit to such data
 (including the full covariance matrix) is provided by a free massive boson 
propagator $G(k) \approx 1/ k^{2} + m^{2}$.
We obtain $m = 590 \pm 30$ MeV, with $\chi^{2}_{dof} = 1.4$.

Since from the above analysis the results at $\beta=5.7$ seem to differ 
significantly from those at $\beta=6.0$ and 
$\beta=6.3$, we have investigated the effect of improving the accuracy of 
the gauge fixing. 
After 300 additional sweeps on each configuration, we repeat our measurements.

For $G(\vec{k}=0, t)$ it turns out that while the central values of the 
individual data points do change significantly, the pattern of the effective 
mass $a m(t)$ is basically the same.
Not surprisingly, 
the best fit is still provided by a simple exponential, with a value for the 
mass parameter which is close to the one obtained before the additional 
sweeps; in other words, to a first approximation the effect of additional 
gauge fixing has been a rescaling of the data points. As far as $G(k)$ is 
concerned, the individual data points are not much affected by 
the additional
gauge fixing.
In fact, the best fit to the data is still provided by a 
free massive boson, but now we obtain $m = 630 \pm 40$ MeV, with 
$\chi^{2}_{dof} = 3.3$.
Moreover, the error bars of some data points have 
become slightly bigger (see Fig.~\ref{fig:fit57}).

Summarizing, we find that the qualitative behavior of the data is not 
changed
by the additional gauge fixing, maintaining the discrepancy with the 
other lattices. 
However,
 since the value of $\chi^{2}_{dof}$ 
in the fit for $G(k)$ has 
increased as well as some individual error bars for the data points,
the additional gauge fixing seems to have increased the ``noise" in the 
evaluation of $G(k)$.   This may be related to the existence of
Gribov copies in our gauge \cite{Sme}.

To go back to the main issue, 
which is the discrepancy between $G(k)$ at 
$\beta=5.7$ and at weaker couplings, our present data give some 
indication that such a discrepancy may be 
due to a lack of scaling at $\beta=5.7$. To illustrate this point we have 
plotted for each data set the quantity $100 \times G(k)/G(k=0)$. Such a 
quantity should not 
depend on $\beta$, as long as one is in the scaling region. On the other 
hand, when comparing $G(k)/G(k=0)$ at different $\beta$ one has an 
uncertainty related to the determination of the horizontal scale, i.e. the 
scale of physical momenta, depending on the values of 
$a^{-1}$. With this {\it caveat} in mind, we show in Figs.~\ref{fig:compa1} 
and \ref{fig:compa2} 
the quantity  
$100 \times G(k)/G(k=0)$ as obtained from our different data sets.
(For the lattice at $\beta=5.7$ we use the data after the additional gauge 
fixing.) 
We again assume 
$a^{-1}=1.2$ GeV at $\beta=5.7$, $a^{-1}=2.1$ GeV at $\beta=6.0$ and 
$a^{-1}=3.2$ GeV at $\beta=6.3$.

\null From the above mentioned 
figures it turns out that in general the data agree  
well,
but significant deviations are observed in the low momentum region; 
in particular, the data at $\beta=5.7$ for momenta up to 
$\approx 650$ MeV 
are rather different from the 
corresponding data at $\beta=6.0$ (see 
Fig.~\ref{fig:compa1}). This may account for the observed discrepancies, 
although further analysis and better 
statistics are needed to clarify  
this issue. 
 
\bigskip
At this point it is worth mentioning that we have checked the stability 
of our fits for $G(k)$ versus the use of continuum and lattice formulas. 
In fact, especially when the the data points correspond to 
lattice momenta of order $a^{-1}$, one 
may wonder whether more accurate fits could be obtained by using lattice 
versions of our formulas. 
One can devise such expressions 
by substituting $k^2$ with $2 / a^{2} \ \sum_{\mu} \ (1 - cos(k_{\mu} a))$ 
in the continuum ones. 

 We have also checked the stability of the results
under change  from covariant to noncovariant fits.
It turns out that when the covariance matrix is well behaved, 
which is what typically happens on our lattices for $G(k)$, the fit is not 
very sensitive to the different definitions for $\chi^{2}$. Moreover, the fits 
are not 
sensitive to the difference between continuum and lattice formulas, not even 
in the higher momentum region. This is shown in Fig.~\ref{fig:testlat}
for the $24^4$ lattice at $\beta=6.3$, where we compare lattice and continuum 
versions of the best fit to an inverse power law, with and without the full 
covariance matrix. 

One gets from the fit to the lattice formula, without covariance matrix, 
$\gamma \approx .69$, and from the fit to the continuum formula, without 
the covariance matrix, $\gamma \approx .61$. These numbers should be compared 
to $\gamma = .68 \pm .08$, which is the result we obtained from the fit 
to the continuum power law which included the full covariance matrix.

\section{Conclusions}

Let us summarize our results.

We have evaluated $G(\vec{k}=0,t)$ and $G (k)$ on four different
lattices at three different values of $\beta$. All the data provide
evidence for dynamical mass generation, in agreement
with previous results for $G(\vec{k}=0,t)$.
\begin{enumerate}
\item For $G(\vec{k}=0,t)$ the data at $\beta=6.0$ and $\beta=6.3$ show 
an effective gluon mass which increases with the time separation, for 
short times, in agreement with previous results. At $\beta=5.7$ 
there is no clear signal for an increasing
effective mass over more than 1 time slice. A more detailed analysis 
of $G(\vec{k}=0,t)$ is not possible, due to statistical difficulties.  

\item The most interesting quantity is the momentum space propagator 
$G (k)$, which we have evaluated for the first time. 
 It is characterized by a clear numerical signal; in particular, 
it turns out that the data in momentum space are
much less correlated than in real space.
At $\beta=6.0$ the propagator $G (k)$ 
up to momenta $\approx $ 1 GeV
can best be described  by the dynamical mass generation mechanism 
$\grave{a} \ la$ Gribov; however, a description by a standard massive 
propagator is not 
ruled out.  At higher momenta the propagator is best described
 by an inverse power law,  which could be interpreted in terms of an anomalous 
dimension. 
The results at $\beta=6.3$ are consistent, but only
the high momentum region is accessible because of the 
small physical size of the lattice.
Thus the 
behavior of the 
propagator for the above values of $\beta$ could be 
summarized by the determination of a 
mass scale $b$ and an anomalous dimension $\gamma$. In spite of the 
uncertainties in the determination of $a^{-1}$, our numerical
values for $\gamma$ are consistent between $\beta=6.0$ and $\beta=6.3$
lattices, when assuming $a^{-1} = 2.1$ GeV at $\beta=6.0$ and
$a^{-1} = 3.2$ GeV at $\beta=6.3$.  For $b$, the results are
consistent between the two different volumes at $\beta=6.0$,
which are the only lattices on which it is determined. \par

On the other hand, at $\beta=5.7$ $G (k)$ is best fit, 
up to momenta $\approx $ 1 GeV,
by a simple massive propagator.
 We argue that 
such a discrepancy may be related to a lack of scaling 
between $\beta=5.7$ and 6.0 for $G(k)$ at low momenta. \par

In some cases, 4-parameter fits to particle $+$ ghost expressions for $G(k)$ 
result in small values of $\chi^2_{dof}$,  
 but as the parameters are typically 
very poorly determined, such fits are not very illuminating. 
\end{enumerate}

In conclusion, the lattice study of gluon correlation functions, in the 
gauge-fixed framework which we have discussed, appears to be technically 
possible, although challenging. In particular, the feasibility of the study 
in momentum space
 is very promising for future applications. 

A very careful analysis of systematic lattice effects 
and statistical errors is necessary. 
At the same time, a study of the gauge dependence 
of the mass scales related to the propagator is also in order.

\bigskip

C.P. wishes to thank D. Zwanziger for many illuminating discussions, 
 Y. Shen for an interesting conversation and the Physics Department of the 
University of Rome ``La Sapienza" for hospitality during some stages of the 
present work.
C.B. was partially supported by the DOE under grant number
DE2FG02-91ER40628, and C.P. and A.S. were partially supported under USDOE 
contract number DE-AC02-76CH00016. C.P. also acknowledges financial support 
from C.N.R. The computing for this project was done at the National
Energy Research Supercomputer Center in part under the
``Grand Challenge'' program and at the San Diego Supercomputer Center.

\vfill

\newpage

\section*{Figure Captions}

\begin{figure}[h] 
\caption{$A_{0} (t)$ vs. $t$ for 
one gauge-fixed configuration on the $16^3  
\times
40$ lattice at $\beta=6.0$.}
\label{fig:test}
\end{figure}

\begin{figure}[h] 
\caption{Effective gluon mass in lattice units vs. $t$ 
on the $16^3 \times 40$ lattice at $\beta=6.0$.}
\label{fig:effmass}
\end{figure}

\begin{figure}[h] 
\caption{Momentum space propagator vs. $k$ in GeV on the $16^3 \times 40$ 
and $24^3 \times 40$ lattices at $\beta=6.0$. 
We assume $a^{-1} = 2.1$ GeV.}
\label{fig:3_4} 
\end{figure}

\begin{figure}[h]
\caption{$G (k)$
vs. $k$ in GeV and 
fit to the form (\protect\ref{eq:prop}) 
 on the $16^3 \times 40$ and $24^3 \times 40$ lattices at $\beta=6.0$.} 
\protect\label{fig:5_7}
\end{figure} 

\begin{figure}[h]
\caption{$G (k)$
vs. $k$ in GeV and fit to an inverse power law
 on the $16^3 \times 40$ and $24^3 \times 40$ lattices at $\beta=6.0$. 
In the latter case, the covariance matrix is not included in the fit.}
\protect\label{fig:pow_uv_1}
\end{figure}

\begin{figure}[h] 
\caption{Effective gluon mass in lattice units  vs. $t$ at $\beta=6.3$.
\protect\phantom{fill in line here}}
\label{fig:2effmass}
\end{figure}

\begin{figure}[h] 
\caption{The best fit of $G(k)$ vs. $k$ in GeV to an inverse power law 
(solid line) on the $24^4$ lattice at $\beta=6.3$}
\label{fig:bello}
\end{figure}

\begin{figure}[h]
\caption{Effective gluon mass in lattice units vs. $t$ on the $16^3 \times 
24$ lattice at $\beta=5.7$.}
\label{fig:effmass57}
\end{figure}  

\begin{figure}[h]
\caption{$G(k)$ vs. $k$ in GeV and fit to a massive boson 
propagator before and after 
additional gauge fixing
 on the $16^3 \times 24$ lattice at $\beta=5.7$}
\label{fig:fit57}
\end{figure}

\begin{figure}[h]
\caption{$100 \times G(k)/G(k=0)$ on the different lattices 
(low momentum region).}
\label{fig:compa1}
\end{figure}

\begin{figure}[h]
\caption{Same as Fig.~\protect\ref{fig:compa1}, 
in a higher momentum region. \protect\phantom{~~~~~~~~~~~~~~~~~~~~~~~}} 
\label{fig:compa2}
\end{figure}

\begin{figure}[h]
\caption{ $G(k)$ vs. $k$ in GeV on the $24^4$ lattice at $\beta=6.3$ 
and the best fits to an inverse power law}
\label{fig:testlat}
\end{figure}

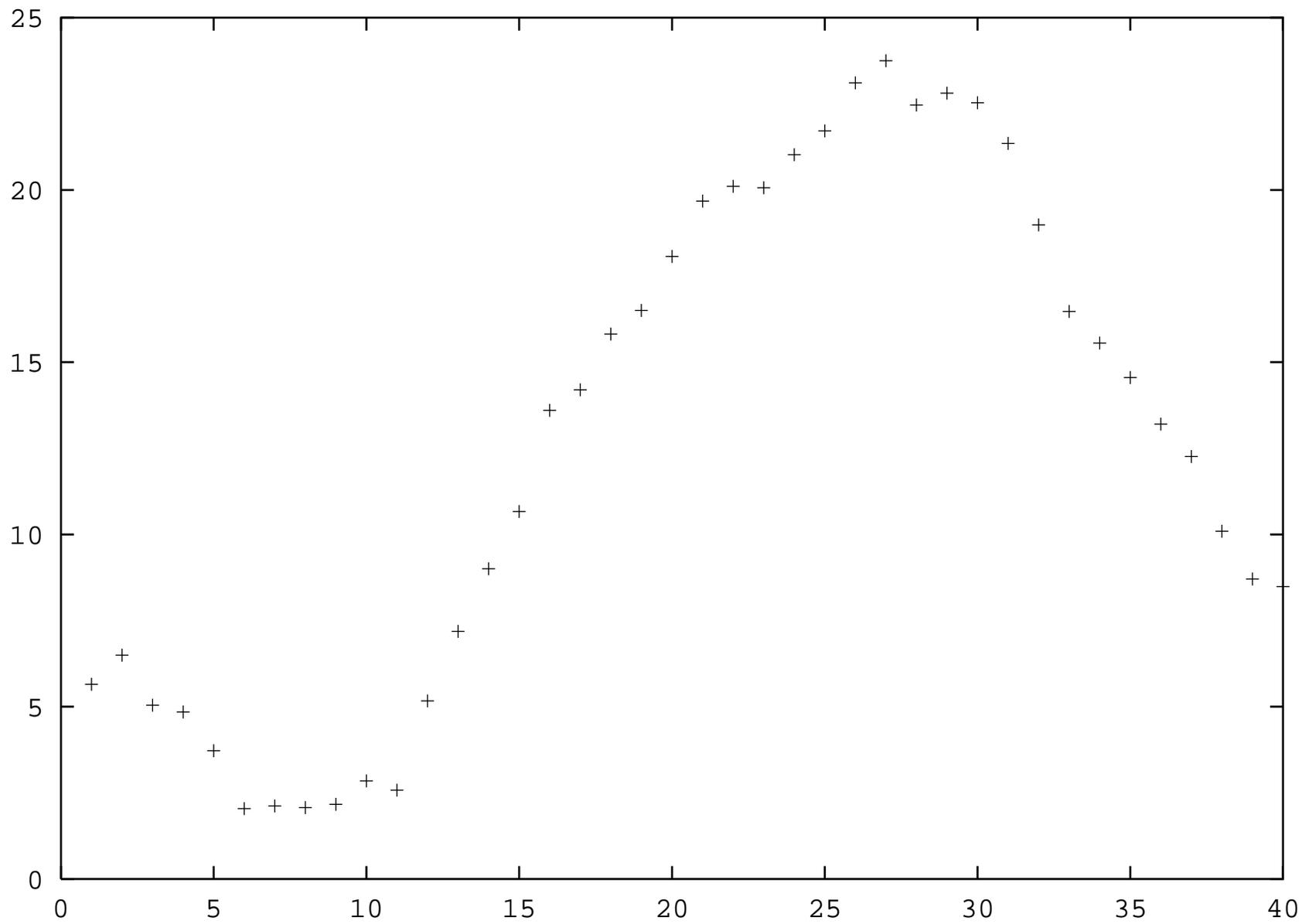
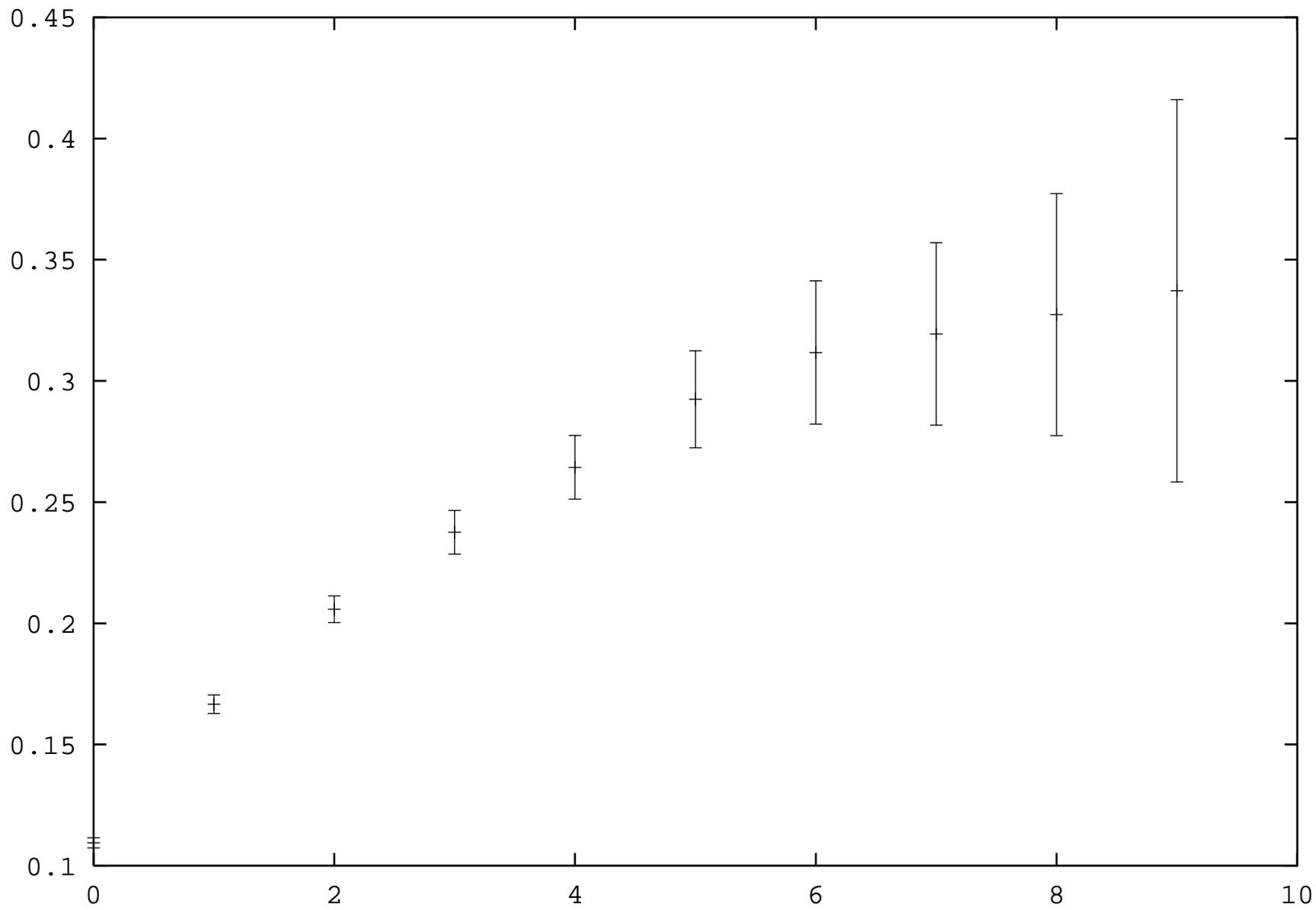
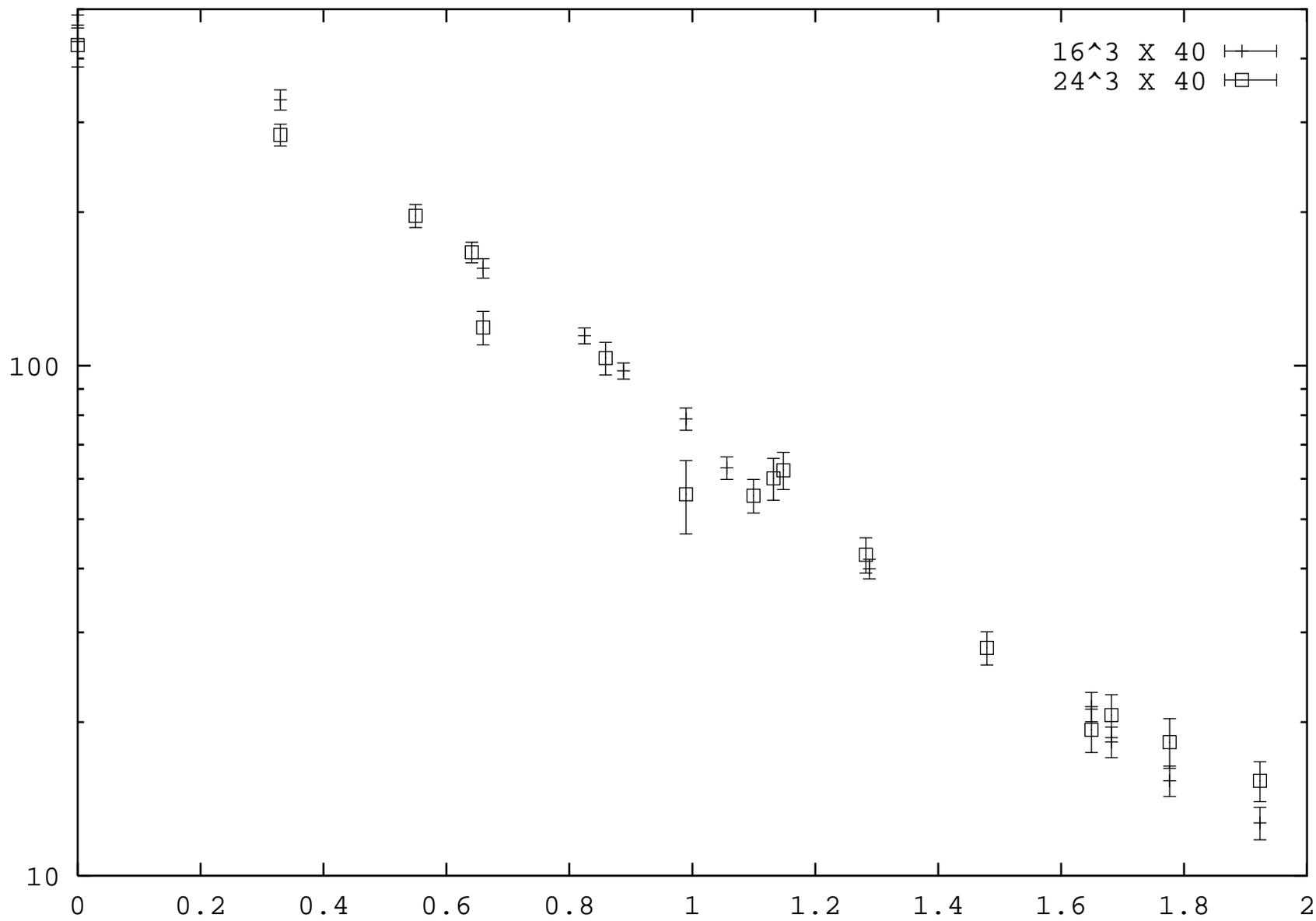
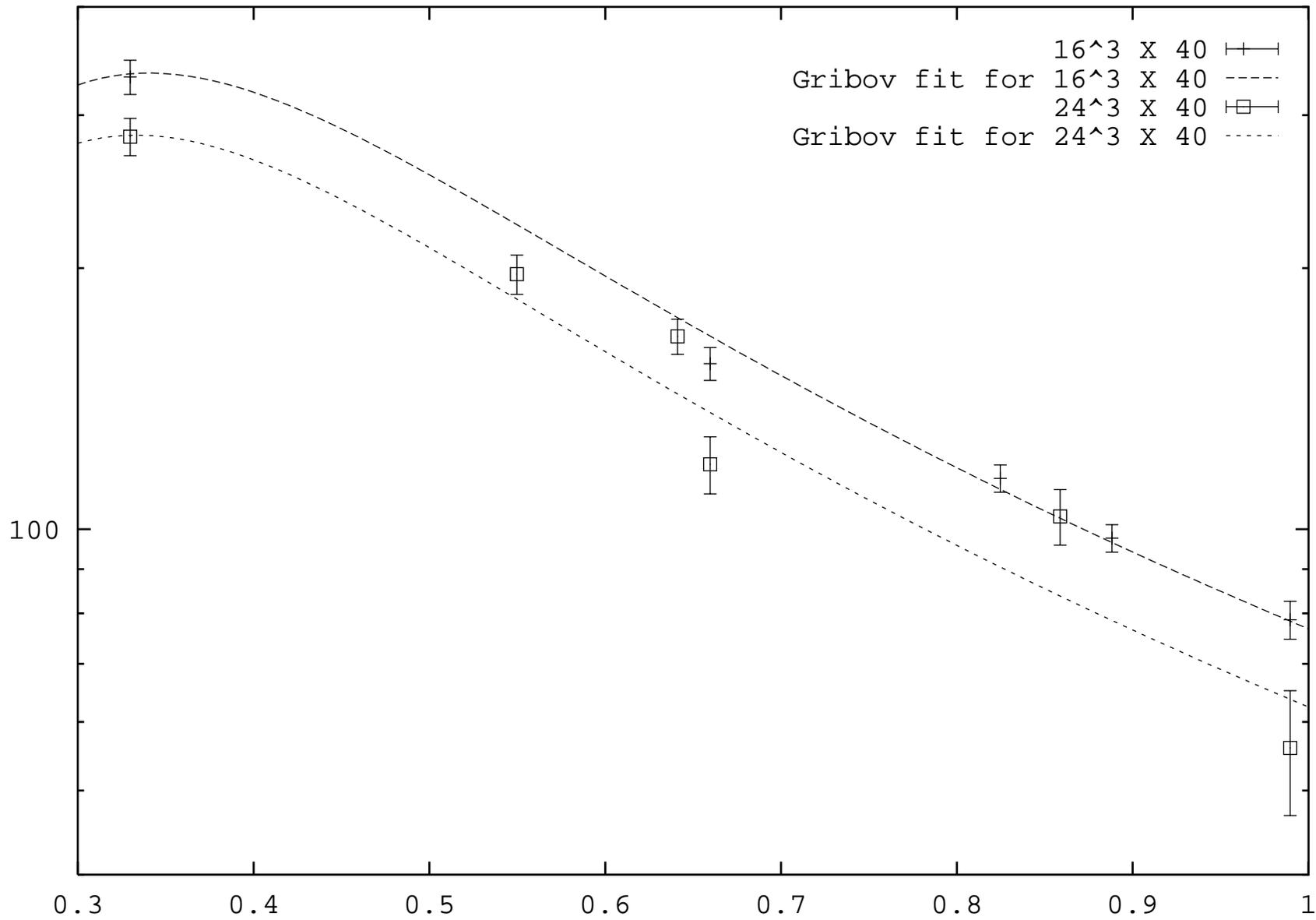
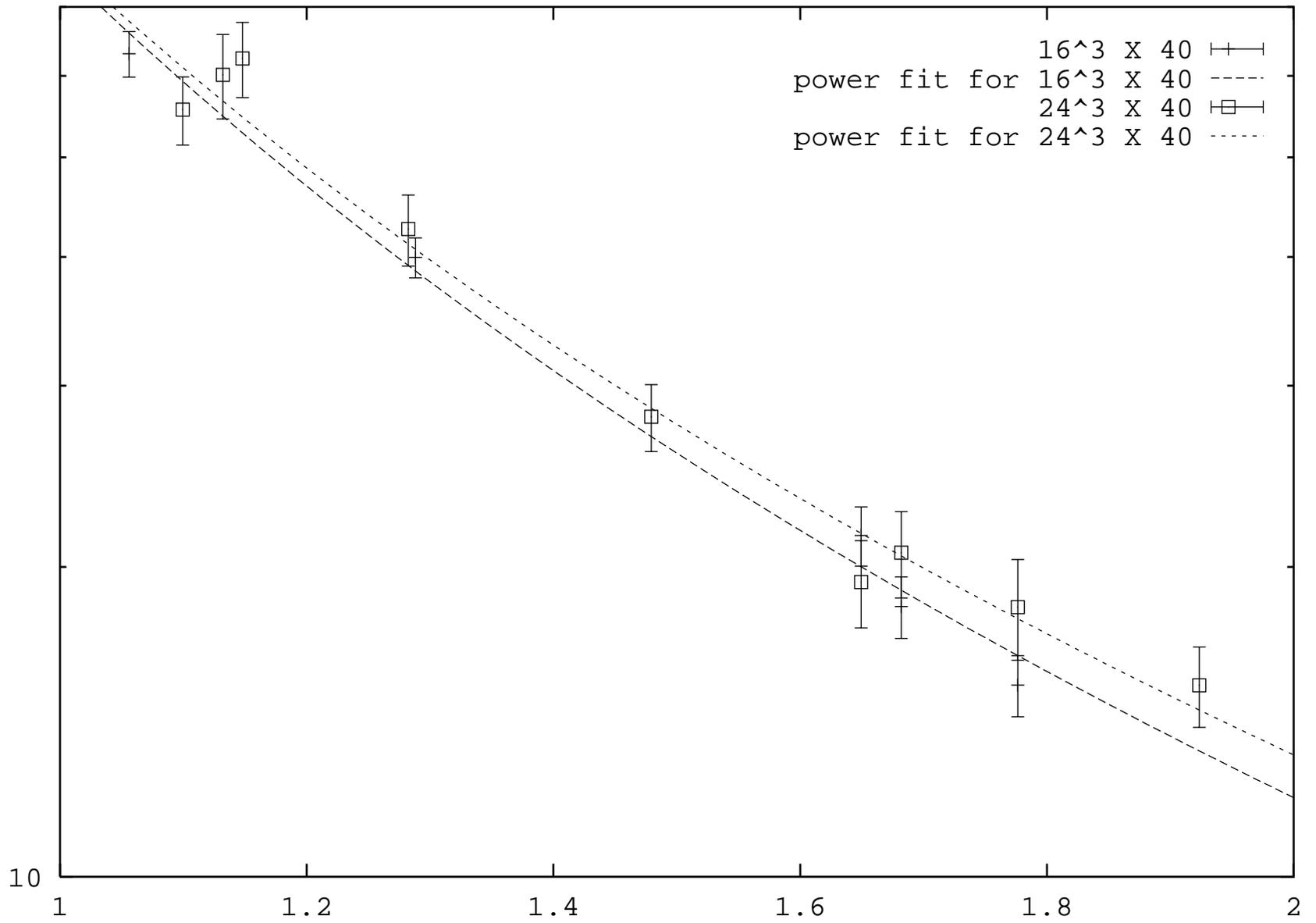
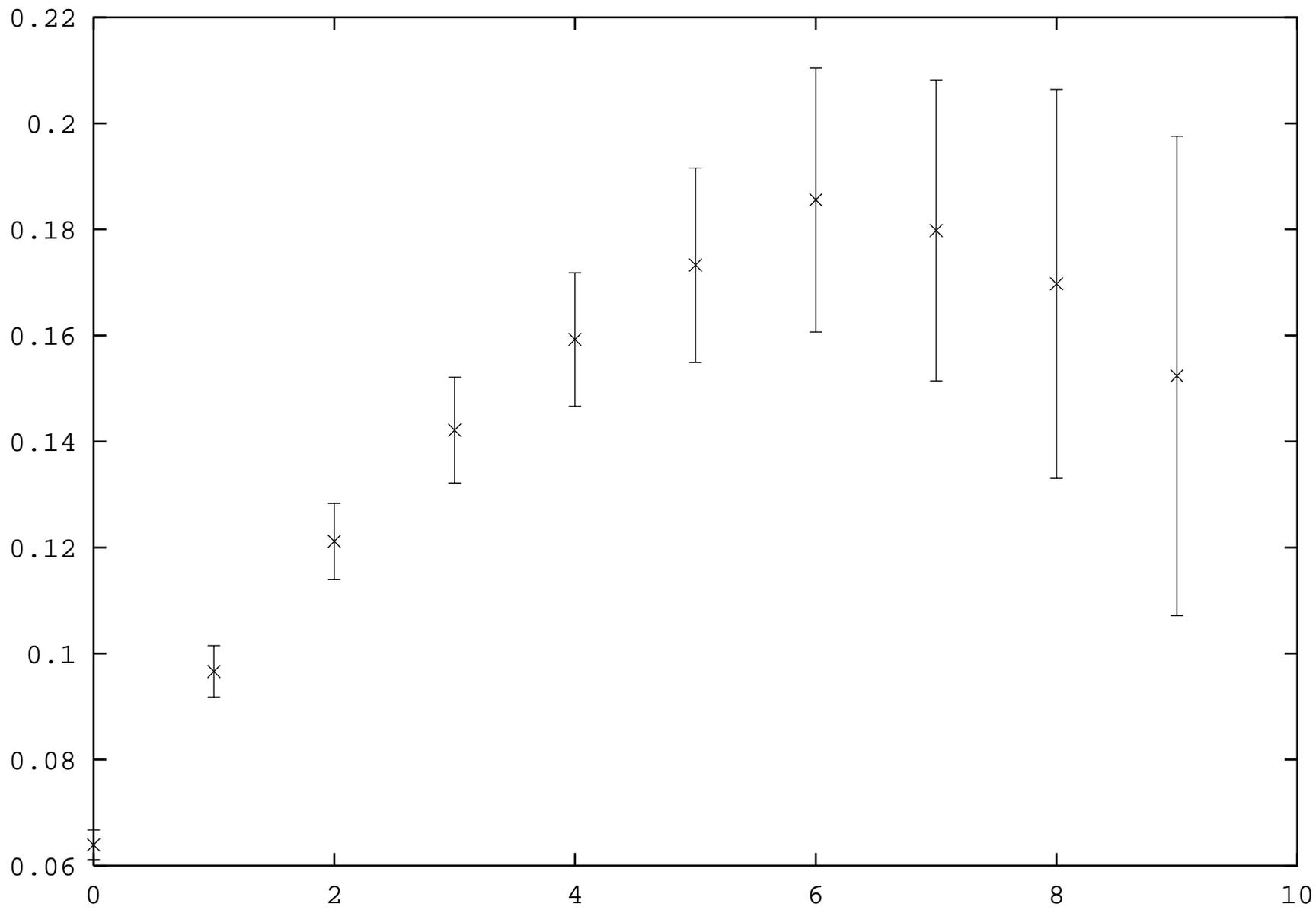
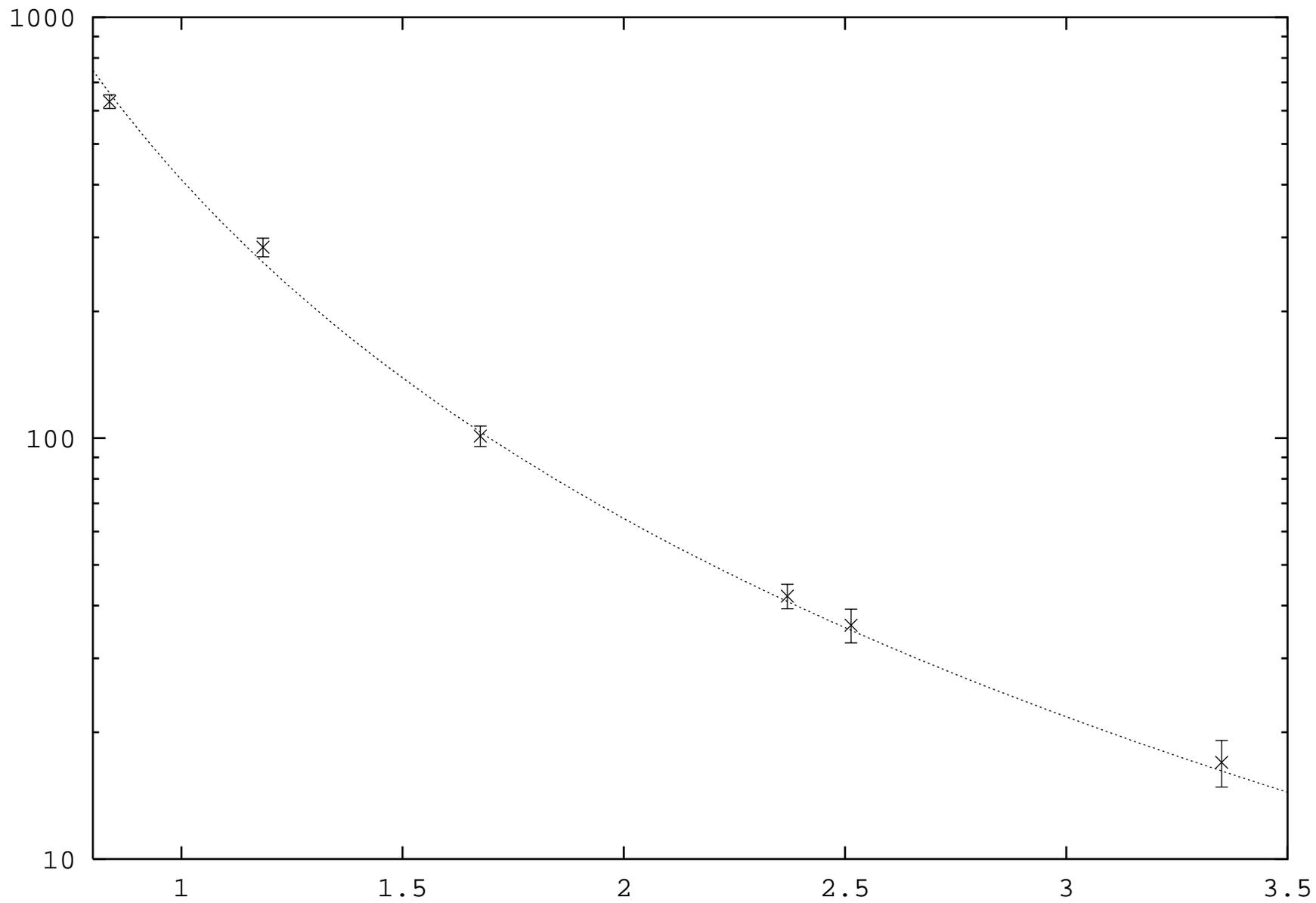
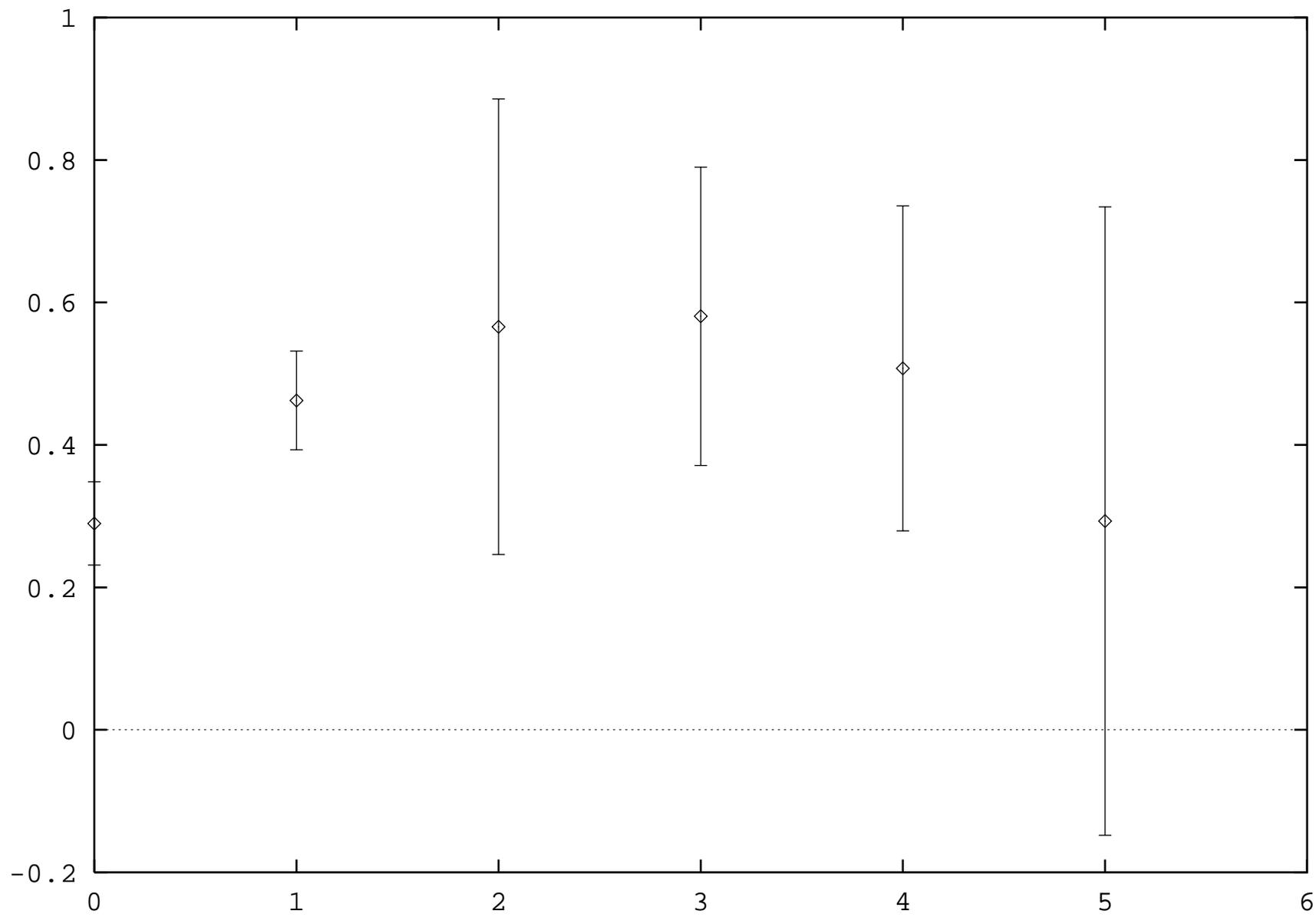
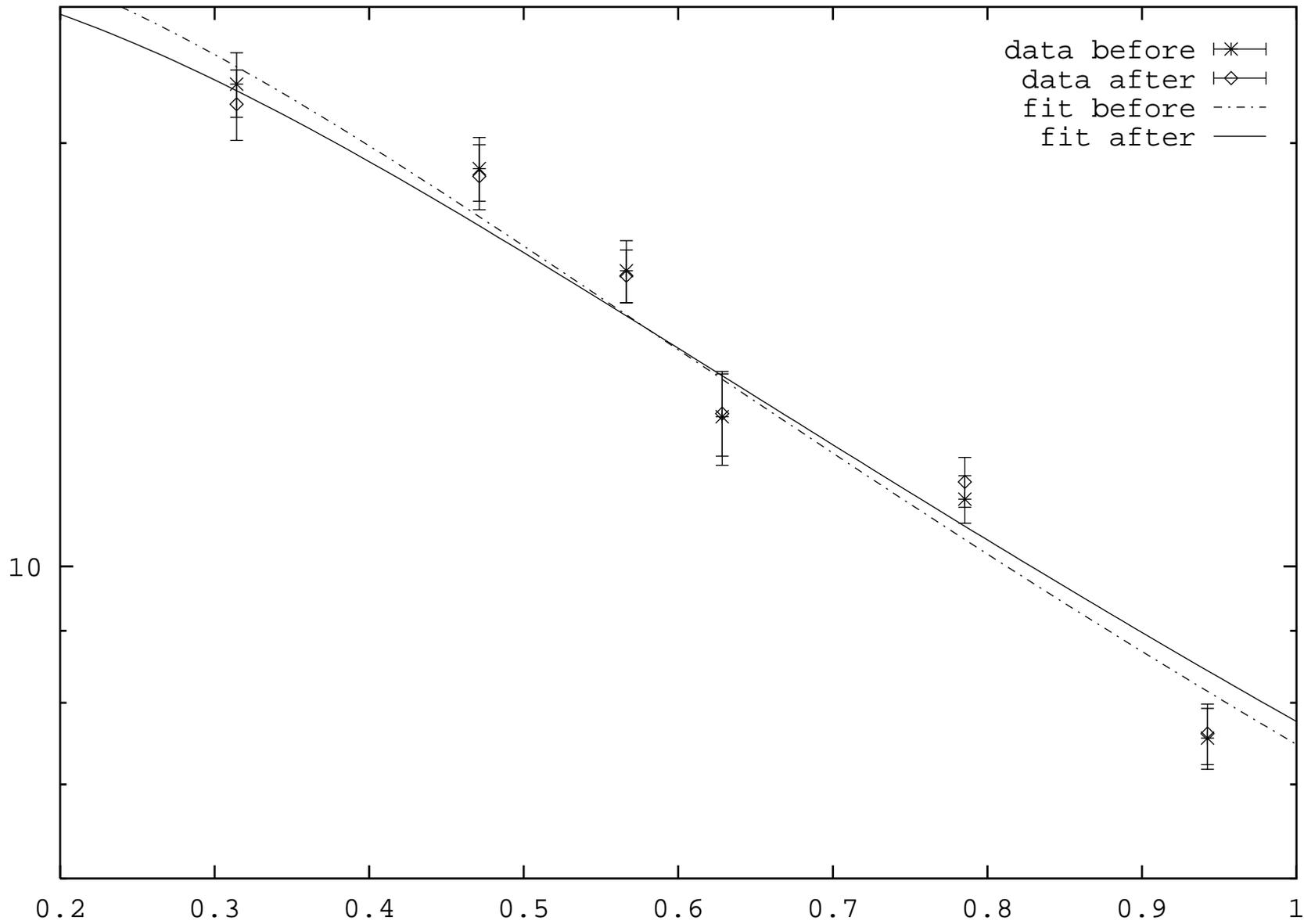
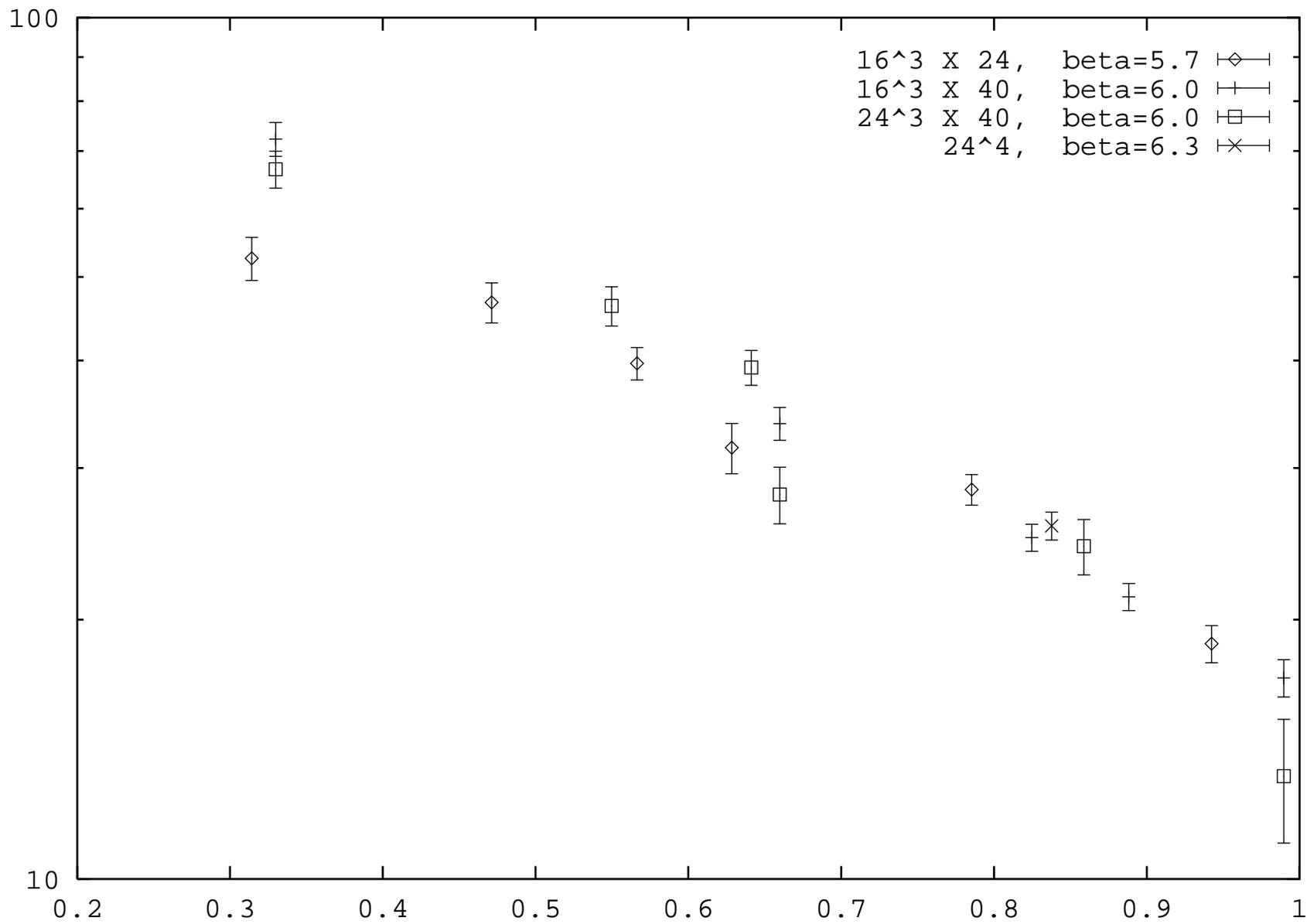
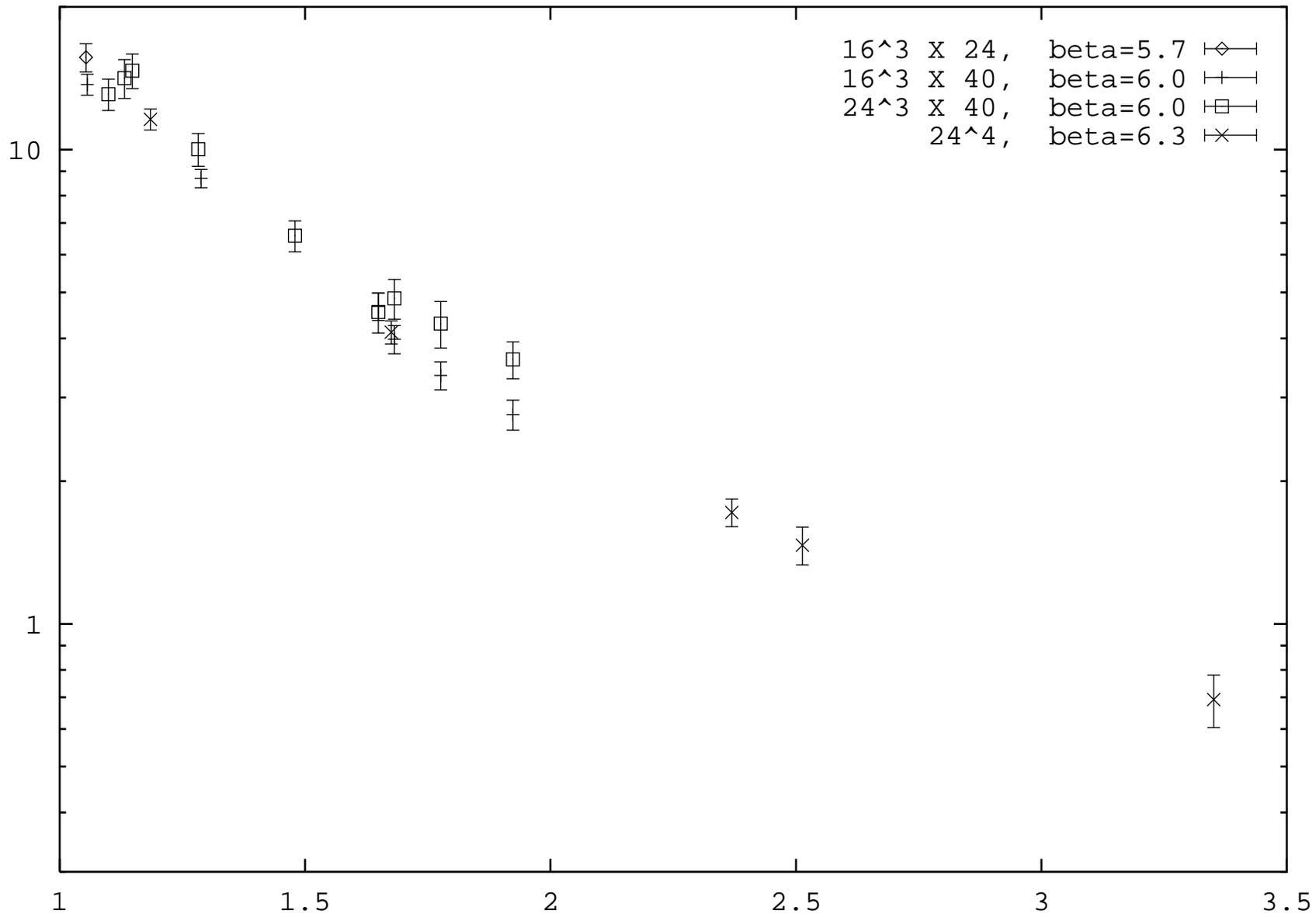
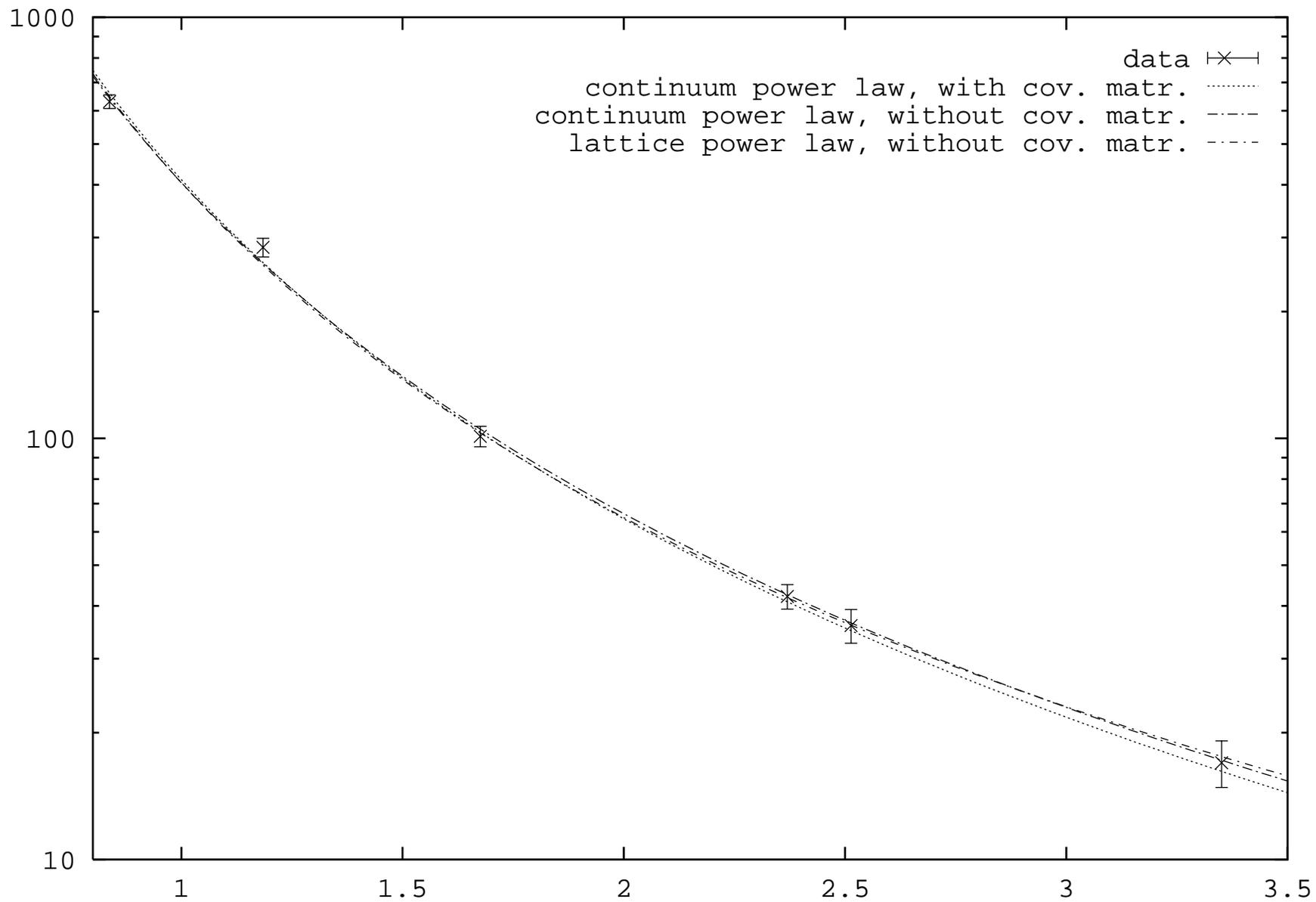
\end{document}